\documentclass[prl,twocolumn,showpacs,preprintnumbers,amsmath,amssymb]{revtex4}
\usepackage{graphicx}
\usepackage{dcolumn}
\usepackage{bm}

\begin{document}

\preprint{APS/123-QED}

\title{Direct Observation of Quadrupolar Excitons\\
in the Heavy-Fermion Superconductor PrOs$_{4}$Sb$_{12}$}

\author{K.~Kuwahara$^1$, K.~Iwasa$^2$, M.~Kohgi$^1$, K.~Kaneko$^3$, N.~Metoki$^{2,3}$, S.~Raymond$^{4,5}$, M.-A.~M{\'e}asson$^4$, J.~Flouquet$^4$, H.~Sugawara$^6$, Y.~Aoki$^1$ and H.~Sato$^1$ }
\affiliation{$^1$Department of Physics, Tokyo Metropolitan University, Tokyo 192-0397, Japan}
\affiliation{$^2$Department of Physics, Tohoku University, Sendai 980-8578, Japan}
\affiliation{$^3$ASRC, Japan Atomic Energy Research Institute, Ibaraki 319-1195, Japan}
\affiliation{$^4$CEA-Grenoble, DRFMC / SPSMS, 38054 Grenoble, France}
\affiliation{$^5$Institut Laue-Langevin, 38042 Grenoble, France}
\affiliation{$^6$Faculty of Integrated Arts and Sciences, The University of Tokushima, Tokushima 770-8502, Japan}

\date{\today}

\begin{abstract}
We report inelastic neutron scattering experiments performed to investigate
the low energy magnetic excitations on single crystals of the heavy-fermion superconductor PrOs$_{4}$Sb$_{12}$.
The observed excitation clearly softens at a wave vector {\it Q}~=~(1,0,0), which is the same as the modulation vector of the field-induced antiferro-quadrupolar ordering, and its intensity at {\it Q}~=~(1,0,0) is smaller than that around the zone center. This result directly evidences that this excitonic behavior is derived mainly from  nonmagnetic quadrupolar interactions.
Furthermore, the narrowing of the linewidths of the excitations below the superconducting transition temperature indicates the close connection between the superconductivity and the excitons.
\end{abstract}



\maketitle


The relation between magnetism and superconductivity in heavy-fermion (HF) systems is one of the most interesting research areas in condensed matter physics. Magnetic fluctuations due to strong Coulomb repulsion and hybridization between f electrons and conduction electrons are believed to be the origin of both the HF behavior and the superconductivity. In general, however, f electrons in systems with high crystal symmetry also have nonmagnetic orbital degrees of freedom and they can interact with conduction electrons through the orbital channels. Therefore, the nonmagnetic orbital fluctuations might also lead to the HF superconductivity. 

The filled skutterudite compounds RT$_{4}$X$_{12}$ (R: rare earth or actinide, T: transition metal, X: pnictogen) crystallize 
in the body-centered cubic  space group $Im${\= 3}, in which R ion is surrounded by a cage of 12 X ions. These compounds show a wide variety of thermal, magnetic and transport properties due to strong $c-f$ mixing effects and a {\lq}rattling{\rq} motion of R ion resulting from the unique crystal structure. 
Among them, PrOs$_{4}$Sb$_{12}$ is the first Pr-based HF superconductor with the superconducting transition temperature $T_{\rm c}$~=~1.85~K \cite{maple02,bauer02}.
The large Sommerfeld coefficient $\gamma \sim$ 350 mJ/mol~K$^2$ of the specific heat and the large jump at  $T_{\rm c}$ ($\Delta C/T_{\rm c}  \sim$ 500 mJ/mol~K$^2$)
suggest that heavy quasiparticles participate in the  superconducting transition.
Specific heat, magnetization and thermal conductivity show double superconducting phase transitions \cite{maple02,tayama03,izawa03}.
 No coherence peak in $1/T_1$ has been found at $T_{\rm c}$ in the NQR experiment  \cite{kotegawa03}. 
 Furthermore, $\mu$SR measurement reveals that the superconducting state breaks the time reversal symmetry \cite{aoki03}. The existence of a field-induced antiferro-quadrupolar (AFQ) ordered phase above $\sim$ 5 T near the superconducting phase naturally suggests that the nonmagnetic quadrupolar degrees of freedom play an important role in this unusual superconductivity \cite{aoki02,kohgi03}. In PrOs$_4$Sb$_{12}$,  4f electrons are well-localized under the cubic crystal field (CF) with $T_h$ symmetry. This is evidenced by dHvA experiment which shows the closeness of the Fermi surface between PrOs$_4$Sb$_{12}$ and LaOs$_4$Sb$_{12}$ \cite{sugawara02} and by inelastic neutron scattering (INS) experiments which show the clear CF excitations \cite{maple02,kuwahara04,goremychkin04}. Compared to other HF superconductors, the well-localized nature of 4f electrons is characteristic to this material. The CF level scheme is well established by elastic \cite{kohgi03} and INS experiments \cite{kuwahara04,goremychkin04} and is composed of the $\Gamma_1$ singlet ground state, the  $\Gamma_4^{(2)}$ triplet first excited states above the CF gap $\Delta$ = 0.7~meV and other excited states located at much larger energy above $\sim$~10 meV. The low-lying $\Gamma_1 - \Gamma_4^{(2)}$ excitations should be responsible for the large $\gamma$ value. Furthermore, the study on Pr(Ru$_{1-x}$Os$_x$)$_4$Sb$_{12}$ series shows that a tiny variation of $\Delta$ leads to a large variation of $\gamma$ \cite{frederick05}.

In this Letter, we report on high resolution INS experiments on single crystals of PrOs$_{4}$Sb$_{12}$, focusing attention on the low-lying singlet-triplet excitations. We found a clear softening of the excitation at the zone boundary {\it Q}~=~(1,0,0) and a peculiar {\it Q}-dependence of the excitation intensity. This result gives the direct evidence for the excitons derived from nonmagnetic quadrupolar interactions between 4f electrons. Furthermore, the temperature dependence of the excitations suggests that the new {\lq}quadrupolar{\rq} excitons are coupled to the HF superconductivity. 

\begin{figure}
\includegraphics[scale=0.28]{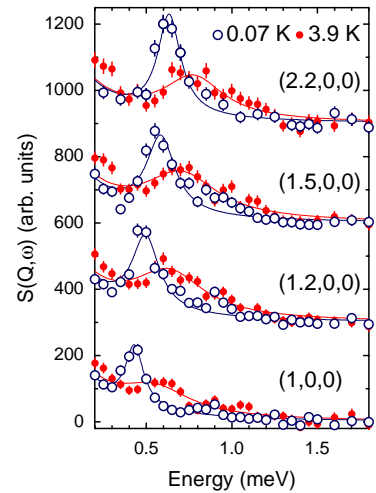}  
\caption{(Color online) Energy spectra of excitations in PrOs$_{4}$Sb$_{12}$ with different {\it Q} vectors along the [1, 0, 0] direction at 0.07 K (open circle) and at 3.9 K (closed circle). Lines are the two Lorentzian fit.}
\label{fig:raw_data1}
\end{figure}

Single crystals of PrOs$_{4}$Sb$_{12}$ were grown by Sb-self-flux method. In order to gain  counting statistics in the INS experiments, we used two single crystals. 
The total mass is about 6.8~g. 
One of them is the same as used for the previous INS experiments \cite{kuwahara04}. Their superconducting transitions 
at $T_{\rm c} \sim 1.8$~K  were confirmed by specific heat measurements performed on small pieces of each crystals. 
Neutron scattering experiments were carried out using the cold neutron three-axis spectrometers IN12 and IN14 located at the Institut Laue-Langevin, Grenoble. 
Data presented in this Letter were exclusively taken on IN14 and are consistent with IN12 preliminary data.
The incident beam was provided by a vertically focused pyrolytic graphite (PG) monochromator. A liquid-nitrogen cooled Be filter placed after the PG monochromator was employed to cut down higher-order contamination of neutrons.
A horizontally focusing PG was used as analyzer. Energy scans at constant {\it Q} along the [1,0,0] direction were performed with the fixed final neutron energy 3.0 meV and with collimation open-60'-open-open.
The energy resolution determined by the incoherent signal is 0.10 meV full width at half maximum (FWHM). The assembly of two single crystals has a total mosaicness of $1.3^{\circ}$, corresponding approximately to the mosaicness of each individual crystal.
The assembly was mounted in a $^3$He-$^4$He dilution refrigerator and a standard $^4$He cryostat with ($hk0$) as horizontal scattering plane. The latter was used especially for measuring excitations in the vicinity of $T_{\rm c}$.
A background was determined by using the neutron energy gain data and was subtracted. 
The intensity of neutron scattering along the [1,0,0] direction I$_{\rm obs}(Q, \omega)$
is proportional to the dynamic structure factor ${\rm S}(Q, \omega) = {\rm S}_{zz}(Q, \omega) + {\rm S}_{yy}(Q, \omega)$ as I$_{\rm obs}(Q, \omega)
\propto 
 A(Q) f^2(Q)
{\rm S}(Q, \omega)$, where $A(Q)$ is the absorption coefficient and $f(Q)$ the magnetic form factor of Pr$^{3+}$ ion. $A(Q)$ was determined by measuring the {\it Q}-dependence of the incoherent scattering intensity under the same geometrical condition for the scattering angle and the sample rotation angle as energy scans in Fig.~\ref{fig:raw_data1}.
By using this experimentally determined absorption coefficient, we got S$(Q, \omega)$.

Figure~\ref{fig:raw_data1} shows the dynamic structure factor S($Q, \omega$) of PrOs$_{4}$Sb$_{12}$ with different {\it Q} vectors along the [1, 0, 0] direction at the lowest temperature 0.07~K and at 3.9~K.
The peaks around 0.4${\sim}$0.8~meV correspond to the ${\Gamma_1}-{\Gamma_4^{(2)}}$  CF transition.
In both superconducting and normal states, the clear softening of the low-lying excitations at the zone boundary {\it Q} = (1, 0, 0) with minimum energies of 0.43~meV and 0.55~meV respectively has been observed. 
Note that the wave vector (1, 0, 0) is the same as that of the field-induced AFQ ordering,
suggesting the importance of the quadrupolar degrees of freedom even in zero magnetic field, which is more directly evidenced by the {\it Q}-dependence of the excitation intensity  as discussed later.
In the superconducting state, the spectra show sharp peaks with linewidths of 0.2 meV, indicating the localized character of 4f electrons.  
In the normal state, on the other hand, the peaks significantly broaden and their positions shift to higher energies.
In Fig.~\ref{fig:raw_data1}, a faint peak seems to be around 1~meV. However, it has not been observed in the IN14 experiment using the $^4$He cryostat. Therefore, we consider that 
the slightly enhanced intensity at about
1 meV appears to be an artifact of the experimental set-up.
The observed S(${\it Q}, \omega$) was fitted to two Lorentzian line shapes, corresponding to the tail of incoherent signal and the peak around 0.4${\sim}$0.8~meV. For all the spectra, the fitting parameters of the incoherent tail are fixed as those determined at the lowest temperature.

\begin{figure}
\includegraphics[scale=0.28]{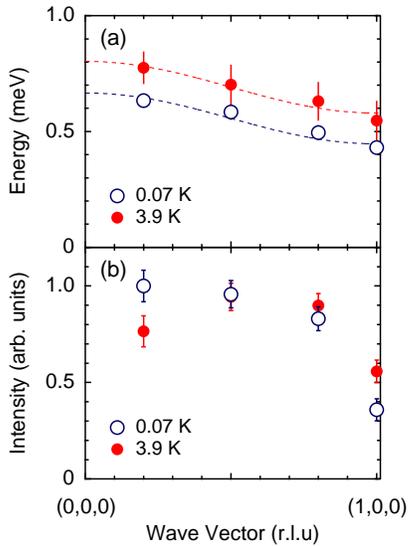}  
\caption{(Color online) (a) Dispersion relations for the peaks around 0.4${\sim}$0.8~meV and
(b) {\it Q}-dependence of the integrated intensity along the [1, 0, 0] direction at 0.07~K  (open circle) and 3.9~K (closed circle) obtained from S($Q, \omega$) in Fig.~\ref{fig:raw_data1}.
In Fig.~\ref{fig2}~(b), the intensity is normalized by that at {\it Q} = (2.2,0,0) for {\it T}~=~0.07~K.
Dashed lines are guides for the eyes}
\label{fig2}
\end{figure}

Figure~\ref{fig2}~(a) is the dispersion curves obtained from the fitting in Fig.~\ref{fig:raw_data1}. 
The energy of the excitations monotonically softens from the zone center to the zone boundary in both the normal and  superconducting states, while the energy decreases uniformly for all {\it Q} vectors with decreasing temperature. 
This behavior seems to be accounted for as magnetic excitons, which have been extensively studied in rare earth metals and compounds for over three decades \cite{jensen91},
but the interactions that modulate the CF levels to create so-called excitons are not necessarily magnetic exchange. 
Indeed, the agreement between the wave vector of the softening and  the modulation vector of the field-induced AFQ ordering  \cite{kohgi03}
most naturally indicate that the interaction between quadrupolar moments ${\bf Q}_i$ remains the principal interaction. 
However, we must also take into account the interaction between magnetic moments ${\bf M}_i$, because the bulk magnetic susceptibility indicates finite antiferromagnetic correlations \cite{tayama03}. 
Therefore, the experimental determination of the nature of the main interaction is a fundamental issue to clarify the mechanism of the superconductivity in PrOs$_4$Sb$_{12}$.

The monotonic decrease of the exciton energy from the zone center to the zone boundary suggests that the nearest-neighbor interaction is dominant in PrOs$_4$Sb$_{12}$.
We, therefore, assume the following effective Hamiltonian of interactions, $H = D_M  \sum_{i,j}{\bf M}_i\cdot{\bf M}_j + D_Q \sum_{i,j}{\bf Q}_i\cdot{\bf Q}_j$ \cite{shiina04_1}, where $D_M$ and $D_Q$ denote the coupling constants of magnetic and quadrupolar interactions, respectively.
The Fourier transforms of the coupling constants in $bcc$ lattice are $D_i({\bf q}) = 8 D_i \prod_{j = x,y,z}\cos(q_j a/2)$ ($i = M, Q$).
If magnetic and quadrupolar interactions coexist in the singlet-triplet system, the exciton dispersion relation within the random phase approximation is written as \cite{shiina04_2}
\begin{equation}
\hbar \omega_{\bf q} = 
\sqrt{( \Delta + 2 |\alpha|^2 D_M({\bf q}) )( \Delta + 2 |\beta|^2 D_Q({\bf q}) ) },
\label{eq:dispersion}
\end{equation}
where $\alpha$ and $\beta$ are the off-diagonal matrix elements of magnetic and quadrupolar moments between  ${\Gamma_1}$ and ${\Gamma_4^{(2)}}$, respectively \cite{matrixelements}. When $D_Q = 0$,  Eq.~(\ref{eq:dispersion}) becomes the usual form of magnetic excitons \cite{jensen91}.
The observed dispersion curves follow the antiferro-case of  Eq.~(\ref{eq:dispersion}), {\it i.e.} $D_M, D_Q > 0$.
As pointed out by Shiina {\it et al.}, the magnetic and nonmagnetic interactions give the same energy dispersion but they give  opposite {\it Q}-dependence of the intensity. Therefore, the intensity analysis is a definitive test that can distinguish the nature of the interaction. The theoretical integrated intensity ${\rm I}({\bf q}) = \int{\rm S}({\bf q},\omega)d\omega$ is given by \cite{shiina04_2}
\begin{equation}
{\rm I}({\bf q}) \propto \sqrt{\frac{\Delta+2 |\beta|^2 D_Q({\bf q})}{\Delta+2 |\alpha|^2 D_M({\bf q})}}.
\label{eq:int}
\end{equation}
When the antiferro-quadrupolar fluctuations give rise to the dispersion of excitations, Eq.~(\ref{eq:int}) predicts that 
${\rm I}({\bf q})$ at the zone boundary, where the $nonmagnetic$ quadrupolar fluctuations are stronger, is smaller than 
${\rm I}({\bf q})$ at the zone center. The opposite behavior comes from the fact that ${\rm I}({\bf q})$ is proportional to the $magnetic$ correlation function.
This conclusion is universal and also does not depend on the details of system.

In Fig.~\ref{fig2}~(b), we plot the {\it Q}-dependence of the integrated intensity determined by subtracting the incoherent signal from S($Q, \omega$) in Fig.~\ref{fig:raw_data1} and summing the remaining counts.
The observed  {\it Q}-dependence of the intensity is quite different from  that of the usual magnetic excitons. The intensity at the zone boundary is smaller than that around the zone center in both normal and superconducting states.
The experimental results agree with Eq.~(\ref{eq:int}) for $|\beta|^2 D_Q \geq |\alpha|^2 D_M$. This result directly evidences that the observed excitonic behavior is  derived mainly from nonmagnetic quadrupolar interactions. Within our knowledge, PrOs$_4$Sb$_{12}$ is the first example of the excitons due to quadrupolar interactions.

\begin{figure}
\includegraphics[scale=0.28]{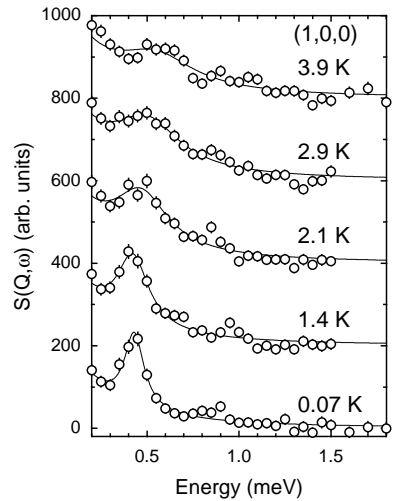}  
\caption{Energy spectra of excitations in PrOs$_{4}$Sb$_{12}$ at {\it Q}~=~(1,0,0) for different temperatures. Lines are the two Lorentzian fit.}
\label{fig:raw_data2}
\end{figure}

\begin{figure}
\includegraphics[scale=0.35]{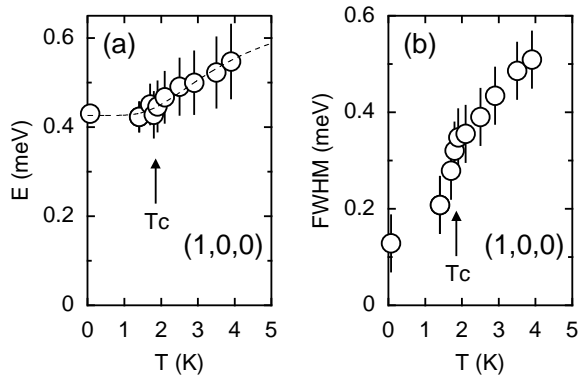}   
\caption{Temperature dependence of (a) the energy and (b) the linewidths of the peaks at {\it Q}~=~(1,0,0).
Dashed line denotes the theoretical curve of the excitons (see text).}
\label{fig:Tdep}
\end{figure}

We also observed the strong temperature dependence of excitation spectra at {\it Q} = (1, 0, 0) as shown in Fig.~\ref{fig:raw_data2}. By the same fitting procedure as in Fig.~\ref{fig:raw_data1}, we plot the temperature dependence of the energy and the linewidths (FWHM) of the peaks  around 0.4${\sim}$0.8~meV in Fig.~\ref{fig:Tdep}~(a) and (b), respectively.
Assuming that the interactions are only quadrupolar, the temperature dependence of the exciton energy is given by multiplying the term of $D_Q({\bf q})$ of Eq.~(\ref{eq:dispersion}) by a temperature renormalization factor $\frac{1-\exp(-\Delta/k_{\rm B} T)}{1 + 3 \exp(-\Delta/k_{\rm B} T)}$, in the same way as usual magnetic excitons \cite{jensen91}. 
The theoretical curve for $\Delta$ = $0.7 \pm 0.1$~meV and $\beta^2 D_Q$ = $0.029 \pm 0.01$~meV approximately reproduces the observed peak shift as shown by the dashed line in Fig.~\ref{fig:Tdep}~(a).
It is obscure whether the energy of the peaks decreases gradually or decreases with a distinct anomaly at $T_{\rm c}$ as decreasing temperature. This ambiguity may come from 
the fact that the superconducting energy gap is comparable  with the CF  gap  $\Delta$ = 0.7~meV.
On the other hand, the linewidths of the peaks exhibit a discontinuous decrease just below $T_{\rm c}$ as shown in Fig.~\ref{fig:Tdep}~(b), suggesting the close connection between the quadrupolar excitons and the superconductivity.
This result agrees with the general observation that for rare earth ions dissolved in superconducting materials,
the CF linewidth reacts to the Cooper pairing \cite{feile81,mesot98}. Exciton based superconductivity theory also predicts that the life time of excitons gets longer below $T_{\rm c}$ \cite{matsumoto04}. 
The novelty of PrOs$_4$Sb$_{12}$ is the exotic feedback between $\Delta$ and $\gamma$ due to the proximity of multipolar instability.
The open underlying questions are if the coherence among the excitons is associated with the entrance in a multipolar state of tiny order parameter and if there is any link of our neutron results with  double structure of the specific heat at $T_{\rm c}$.

The usual BCS superconductor PrRu$_4$Sb$_{12}$ with normal effective mass \cite{frederick05,takeda00,abe02} also has the singlet-triplet CF level scheme but the CF gap between $\Gamma_1$ and $\Gamma_4^{(2)}$ 5.26~meV, which has been determined by our recent INS experiments on a polycrystal sample \cite{PrRuSb_INS}, is much larger than $\Delta$ of PrOs$_4$Sb$_{12}$. 
This fact clearly indicates that the smallness of $\Delta$, namely the existence of some degrees of freedom in low energy region, is the necessary condition for the mass enhancement. 
This is consistent with a mechanism of the mass enhancement through the $c-f$  magnetic exchange in the singlet ground state system~\cite{fulde83}. In addition, when 4f electrons have magnetic degrees of freedom in such low energy region, the local magnetic fluctuations may be expected at low temperatures. In such a case,
the resulting magnetic quasi-elastic scattering is usually detected in INS experiments as in Ce-based HF compounds.
In PrOs$_4$Sb$_{12}$, however, quasi-elastic scattering has not been detected at least above 0.2~meV as shown in Figs.~\ref{fig:raw_data1} and \ref{fig:raw_data2}. This result suggests other nonmagnetic mechanism rather than that due to magnetic fluctuations. On the other hand, 
the significant broadening of spectral shape in the normal state clearly suggests the importance of the $c-f$ interaction. 
Combining these facts, the mass enhancement as well as the superconductivity in PrOs$_4$Sb$_{12}$  may originate from the strong $c-f$ interaction through the quadrupolar channels.

In summary, new {\lq}quadrupolar{\rq} excitons, which are CF excitations modulated by quadrupolar interactions, are realized in PrOs$_4$Sb$_{12}$.
A clear softening of this excitation occur at {\it Q}~=~(1,0,0). On cooling, all the exciton branch softens.
Moreover the tendency of the narrowing of linewidths below  $T_{\rm c}$ suggests that the quadrupolar excitons are related to the HF superconductivity and vice versa. 

We wish to thank R. Shiina, O. Sakai, M. Matsumoto and M. Koga for helpful discussions and the cryogenic staffs of  the ILL for technical assistance.
This work was supported by the Grant-in-Aid for Scientific Research on the Priority Area {\lq}{\lq}Skutterudites{\rq}{\rq} from MEXT of Japan.

\end{document}